\documentclass[10pt,conference]{IEEEtran}
\usepackage{amssymb}
\usepackage{amsfonts,amsmath,graphicx,amssymb,algorithm, algorithmic,graphs}
\newcommand{\bl}{\boldsymbol}
\newcommand{\ml}{\mathcal}
\newcommand{\diff}{{d\over dt}}
\newtheorem{lemma}{Lemma}
\newtheorem{theorem}{Theorem}
\newtheorem{corollary}{Corollary}

\begin{document}

\title{Greedy Maximal Scheduling in Wireless Networks}
\author{\authorblockN{Qiao Li}
\authorblockA{qiaoli@cmu.edu\\
Department of Electrical and Computer Engineering \\
Carnegie Mellon University \\
5000 Forbes Ave., Pittsburgh, PA 15213}
\and
\authorblockN{Rohit Negi}
\authorblockA{negi@ece.cmu.edu\\
Department of Electrical and Computer Engineering \\
Carnegie Mellon University \\
5000 Forbes Ave., Pittsburgh, PA 15213}}

\maketitle

\begin{abstract}
In this paper we consider greedy scheduling algorithms in wireless
networks, i.e., the schedules are computed by adding links greedily
based on some priority vector. Two special cases are considered: 1)
Longest Queue First (LQF) scheduling, where the priorities are
computed using queue lengths, and 2) Static Priority (SP)
scheduling, where the priorities are pre-assigned. We first propose
a closed-form lower bound stability region for LQF scheduling, and
discuss the tightness result in some scenarios. We then propose an
lower bound stability region for SP scheduling with multiple
priority vectors, as well as a heuristic priority assignment
algorithm, which is related to the well-known
Expectation-Maximization (EM) algorithm. The performance gain of the
proposed heuristic algorithm is finally confirmed by simulations.
\end{abstract}

\section{Introduction}
\label{sec_intro}

Optimal scheduling in wireless networks is, in general, an
NP-complete problem. Essentially, in order to achieve the optimal
stability region, one either needs to solve an NP-complete problem
in each time slot (see the max-weight scheduling in
\cite{tassiulas92}), or approach the optimal solution gradually over
time slots (see the linear random scheduler in \cite{tassiulas98},
and the CSMA-type scheduler in \cite{jiang08}), thereby reducing the
computation complexity in each time slot by amortization over a
relatively long period of time. The first case is clearly not
practical, due to the high computation complexity. For the second
case, the delay performance may be quite bad because the queue
lengths can become very large even with low traffic, since,
intuitively, it takes an exponential number of time slots to
converge to a (near) optimal solution.

Recently, there have been significant research activities on
sub-optimal scheduling algorithms with provable performance
guarantees. In \cite{chaporkar08}, maximal scheduling was proposed
as a low (linear) complexity algorithm for wireless networks. In
maximal scheduling, the only constraint is that the scheduled set of
links is \emph{maximal}, i.e., no more link can be added to the
schedule without violating the interference constraint. It has been
shown \cite{chaporkar08} that maximal scheduling can achieve a
constant approximation ratio in typical wireless networks, which is
the fraction of the optimal stability region that can be stabilized
by maximal schedulers. Further, the delay performance of maximal
scheduling is quite good under light traffic. However, since the
class of maximal schedulers is broad, the worst case performance
guarantee of maximal scheduling, in the form of a lower bound
stability region, is pessimistic \cite{chaporkar08}. Thus, it is
necessary to consider specific maximal schedulers for improved
performance guarantees.

In this paper we consider two specific types of maximal schedulers:
Longest Queue First (LQF) scheduling and Static Priority (SP)
scheduling. In LQF scheduling, the schedule is computed by queue
length based priorities, i.e., links are added according to their
queue lengths, from the longest to the shortest, and a link with
non-empty queue is added to the schedule whenever there is no
conflict. In the literature, it has been shown that LQF scheduling
is optimal if the network satisfies the ``local-pooling condition''
\cite{dimakis05}, which is a function of the network topology.
Later, it was generalized to the notion of ``local-pooling factor''
\cite{joo08}, which was shown to be equal to the approximation
ratio. However, to the best of the authors' knowledge, it is hard to
specify the stability region of LQF scheduling. That is, given a
vector of packet arrival rates, it is difficult to predict whether
that rate can be supported by LQF scheduling, since checking the
``local-pooling condition'' requires complexity exponential in the
network size $n$. This is particularly inconvenient for cross-layer
optimization, where one needs to allocate the link rates
efficiently, subject to stability constraints under LQF scheduling.
In this paper, we propose a closed-form lower bound stability region
for LQF scheduling, which is further shown to be tight in some
scenarios. Further, we propose a fast (linear complexity) checking
algorithm, which can decide whether a given arrival rate vector is
inside the lower bound stability region.

We next consider SP scheduling, where the only difference from LQF
scheduling is that the links to be added to a schedule are
considered following (pre-computed) static priorities, instead of
queue lengths. Thus, the implementation is simpler than LQF
scheduling, where the changes in queue lengths often generate a
considerable amount of messages to be exchanged across the network.
Further, SP scheduling has comparable performance to LQF scheduling.
For example, in \cite{li08} and \cite{li09} we have shown that the
stability region of SP scheduling with a single priority can achieve
the same lower bound stability region of LQF scheduling, and with
$n+1$ priority vectors we can achieve the optimal stability region.
In this paper, we try to analyze the performance of SP scheduling
with an arbitrary number $K$ of priority vectors. We first formulate
a lower bound stability region for SP scheduling with $K$ priority
vectors, Next, we propose a heuristic priority assignment algorithm,
which assigns two priority vectors using an Expectation-Maximization
(EM) type algorithm. This algorithm generalizes easily to $K$
priorities. Finally, we demonstrate the performance gain of the SP
scheduling through simulations.

The organization of this paper is as follows: In Section
\ref{sec_system} we describe the queueing network model, In Section
\ref{sec_lqf} we consider the performance of LQF scheduling. Section
\ref{sec_sp} analyzes SP scheduling with multiple priorities,
Section \ref{sec_simulation} demonstrate the simulation results.
Finally, Section \ref{sec_conclusion} concludes this paper.

\section{System Model}
\label{sec_system}

In this section we introduce the system model, which is standard in
the literature.

\subsection{Network Topology and Priority Vector}

We consider the scheduling problem at the MAC layer of a wireless
network, where the network topology is modeled as a conflict graph
$\ml G=(\ml V, \ml E)$. Here, $\ml V$ is the set of $n$ links, and
$\ml E$ is the set of pairwise conflicts, i.e., $(i,j)\in \ml E$
implies that links $i$ and $j$ are not allowed to transmit
simultaneously, due to the strong interference that one link can
cause to the other. Note that this model is extensively used in the
literature, and is suitable to model various physical layer
constraints. For example, in Bluetooth or FH-CDMA networks, the only
constraint is that a node can not both transmit and receive
simultaneously. Thus, two links $(i,j)\in \ml E$ if and only if they
share a common transceiver in the network. As another example, the
ubiquitous 802.11 Distributed Coordination Function (DCF) implies
that two links within two hops can not transmit together, due to the
exchange of RTS/CTS messages. Therefore, two links $(i,j)\in \ml E$
if and only if they are within two-hop distance of each other.

For each link $i\in\ml V$, we define its neighborhood as $\ml
N_i=\{j\in\ml V: (i,j)\in\ml E\}$. We next introduce priority
vectors, which are used later in the scheduling algorithm. A
priority vector $\bl p$ is a $n \times 1$ vector which corresponds
to a permutation of the vector $(1, 2, \ldots, n)^T$. Link $i$ is
said to have lower priority than link $j$ if $p_i>p_j$. Thus, 1 is
the highest priority, and $n$ is the lowest priority. Given a
priority vector $\bl p$, we define a priority weighted graph
incidence matrix $P$, such that $P_{ii}=1$, and $P_{ij}={\bf
1}_{\{j\in\ml N_i\}}{\bf 1}_{\{p_i>p_j\}}$, where ${\bf
1}_{\{\cdot\}}$ is the indicator function, i.e., ${\bf
1}_{\{\text{true}\}}=1$ and ${\bf 1}_{\{\text{false}\}}=0$. Thus,
$P_{ij}=1$ if and only if link $j$ is a higher priority neighbor of
link $i$.

\subsection{Queueing Network}

We assume that time is slotted, and associate each link $i$ in the
network with an external source $A_i(t)$, which is the cumulative
packet arrival during the first $t$ time slots. $\bl A(t)$ is the
vectors of $A_i(t)$. The only constraints on the arrival process are
that 1) it is uniformly bounded in each time slot, i.e., there
exists a positive constant $0<A_{\max}<\infty$, such that for all
$t>0$,
\begin{equation}
A_i(t)-A_i(t-1)\leq A_{\max}, \ \forall i\in\ml V
\end{equation}
and that 2) the arrival processes are subject to Strong Law of Large
Numbers (SLLN), i.e., with probability 1 (w.p.1), we have
\begin{equation}
\lim_{t\rightarrow\infty}\bl A(t)/t=\bl a
\end{equation}
where $\bl a$ is the average arrival rate vector. Note that this
assumption on $\bl A(t)$ is quite mild, since it allows the
processes $\bl A(t)$ to be correlated across time slots as well as
across different links. Thus, it is suited to model practical packet
sources, which are often subject to non-ergodic and correlating
upper-layer mechanisms, such as routing and congestion control.

The queueing equation of the network is as the following
\begin{equation}
\bl Q(t)=\bl Q(0)+\bl A(t)-\bl D(t)
\end{equation}
where $\bl Q(t)$ is the queue length vector at time slot $t$, and
$\bl D(t)$ is the cumulative departure vector during the first $t$
time slots. The departure vector at time slot $t$, which is denoted
as $\Delta \bl D(t)\doteq \bl D(t)-\bl D(t-1)$, must correspond to
an independent set in the conflict graph $\ml G$, so that, no packet
contention happen.

We assume that the following scheduling produced departure vector
$\Delta\bl D(t)$: In each time slot, the scheduler (either
centralized or distributed) considers the links according to the
sequence specified by a priority vector $\bl p(t)$, where $p_i(t)$
is the priority of link $i$ during time slot $t$. Thus, a link with
higher priority is always considered before the links with lower
priorities. A link $i$ under consideration is scheduled if and only
if 1) it has nonempty queue and that 2) when link $i$ is considered,
none of the links in its neighborhood $\ml N_i$ have been scheduled.
Note that, any greedy scheduler can be modeled in this way, with a
proper choice of priority vector $\bl p(t)$. Specifically, LQF
scheduling corresponds to computing $\bl p(t)$ by sorting the queue
length vector $\bl Q(t-1)$, and SP scheduling corresponds to
choosing a static $\bl p(t)$ sequence which are pre-computed
priority vectors.

\subsection{Stability Region}
\label{sec_stability_region}

The performance of a scheduler is evaluated by its stability region,
which is defined as the set of arrival rate vectors $\bl a$ such
that \emph{any arrival process} with average rate $\bl a$ is stable
under the scheduler. We define stability as \emph{rate stability}
\cite{dai00}, i.e., $\lim_{t\rightarrow\infty}\bl D(t)/t=\bl a$
w.p.1. For specific stability regions, consider the following
region: $\ml A_{\text{maximal}}=\{\bl a\in \mathbb{R}^n_+:
a_i+\sum_{j\in\ml N_i}a_j\leq 1, \forall i\in\ml V\}$. That is, the
sum arrival rate in any link's neighborhood is no larger than 1. In
\cite{chaporkar08}, it has been shown that $\ml A_{\text{maximal}}$
can be stabilized by maximal scheduling. As another example, an SP
scheduler with priority $\bl p$ can achieve the stability region
\cite{li08} $\ml A_{\bl p}=\{\bl a\in \mathbb{R}^n_+: \|P\bl
a\|_{\infty}\leq 1\}$, where $\|\cdot\|_\infty$ is the infinity
norm. We next illustrate these concepts with an example.

\begin{figure}[t]
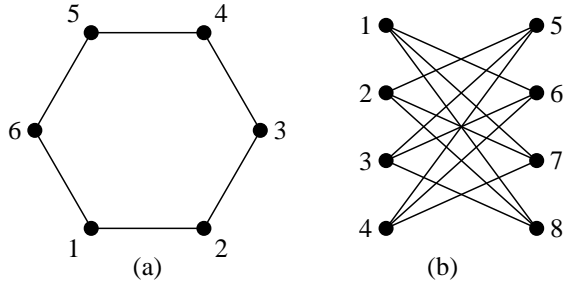

\begin{center}
\begin{tabular}{c c}
\begin{graph}(3.5,3.5)(0,0)
\roundnode{L1}(1, 0.2) \roundnode{L2}(2.5, 0.2) \roundnode{L3}(3.25,
1.5) \roundnode{L4}(2.5, 2.8) \roundnode{L5}(1, 2.8)
\roundnode{L6}(0.25, 1.5) \edge{L1}{L2} \edge{L2}{L3} \edge{L3}{L4}
\edge{L4}{L5} \edge{L5}{L6} \edge{L6}{L1} \autonodetext{L1}[sw]{1}
\autonodetext{L2}[se]{2} \autonodetext{L3}[e]{3}
\autonodetext{L4}[ne]{4} \autonodetext{L5}[nw]{5}
\autonodetext{L6}[w]{6}
\end{graph}
&
\begin{graph}(3.5,3.5)(0,0)
\roundnode{L1}(1, 2.9) \roundnode{L2}(1, 2) \roundnode{L3}(1, 1.1)
\roundnode{L4}(1, 0.2) \roundnode{L5}(3, 2.9) \roundnode{L6}(3, 2)
\roundnode{L7}(3, 1.1) \roundnode{L8}(3, .2) \autonodetext{L1}[w]{1}
\autonodetext{L2}[w]{2} \autonodetext{L3}[w]{3}
\autonodetext{L4}[w]{4} \autonodetext{L5}[e]{5}
\autonodetext{L6}[e]{6} \autonodetext{L7}[e]{7}
\autonodetext{L8}[e]{8} \edge{L1}{L6} \edge{L1}{L7} \edge{L1}{L8}
\edge{L2}{L7} \edge{L2}{L8} \edge{L2}{L5} \edge{L3}{L8}
\edge{L3}{L5} \edge{L3}{L6} \edge{L4}{L5} \edge{L4}{L6}
\edge{L4}{L7}
\end{graph}
\\
(a)&(b)
\end{tabular}
\end{center}
\caption{\label{fig:sample}(a) is a conflict graph consisting of 6
links, and (b) shows an incomplete bipartite graph.}
\end{figure}

\emph{Example:} Consider the 6-node conflict graph in Fig.
\ref{fig:sample}(a), and assume that the priority is $\bl
p=(1,2,3,4,5,6)^T$. Thus, link 1 has the highest priority 1, and
link 6 has the lowest priority 6. For link 1, its neighborhood is
$\ml N_1=\{2, 6\}$. Similarly, we have $\ml N_2=\{1, 3\}$. The graph
incidence matrix $P$ associated with $\bl p$ is as follows:
$P_{ii}=1, 1\leq i\leq 6$, $P_{i(i-1)}=1, 2\leq i\leq 6$,
$P_{61}=1$, and 0 at all the other entries. Let $\bl a=(0.3, 0.4,
0.3, 0.4, 0.3, 0.4)^T$ be an arrival rate vector. Clearly $\bl
a\not\in \ml A_{\text{maximal}}$, since $a_1+\sum_{j\in \ml
N_1}a_j=.3+.4+.4=1.1>1$. But we have $\bl a\in \ml A_{\bl p}$, since
$\|P\bl a\|_\infty=\|(.3, .7, .7, .7, .7, 1)^T\|_\infty=1$.

It is in general difficult to guarantee stability in a network,
since the result has to hold over all arrival processes with the
same average rate. However, characterizing the stability region is
important in some applications, such as in cross-layer design, where
the resources need to be allocated subject to the constraint of
network stability. In next section we formulate a lower bound
stability region of LQF scheduling.

\section{Stability Region of LQF Scheduling}
\label{sec_lqf}

Section \ref{sec_stability_region} argued the importance of
stability region. For LQF scheduling, although the approximation
ratio of LQF scheduling is well-known, and is equal to the
``local-pooling ratio'' of the network \cite{joo08}, the stability
region of LQF scheduling is hard to describe. That is, given an
arrival rate vector $\bl a$, it is difficult to predict that the
network is stable under LQF scheduling without solving a problem of
exponential complexity in $n$. Our previous work \cite{li08} on SP
scheduling allows us to propose the following lower bound stability
region of LQF scheduling: $\ml A_{\text{LQF}}=\{\bl
a\in\mathbb{R}^n_+: \min_{P\in\ml P}\|P\bl a\|_\infty<1\}$, where
$\ml P$ is the set of $P$ matrices associated with $n!$ priority
vectors. We remark that although $\bl P$ has $n!$ matrices, we show
later in the section that optimization involved can be solved with
linear complexity. The stability result is shown in the following
theorem.

\begin{theorem}
The network is rate stable under LQF scheduling for any $\bl a\in
\ml A_{\text{LQF}}$.
\end{theorem}

We first briefly describe the intuition behind the proof. Note that
in LQF scheduling, links with longer queues are always considered
before the links with shorter queues. So, queues that grow are
approximately equal. Thus, if there is a set of links $\ml V_0$
which are the longest, LQF will guarantee that, in each time slot,
the schedule is at least maximal when restricted to the links in
$\ml V_0$. This, together with the fact that $\|P\bl a\|_\infty<1$
for some $P\in \ml P$, guarantees that \emph{some} queue in $\ml
V_0$ is decreasing, which implies that the max queue length is also
decreasing.

\begin{IEEEproof}
In the following proof we will use the technique of fluid limits
\cite{dai00} to prove rate stability. For a brief introduction about
the derivation of fluid limits, please see Appendix.

Let a fluid limit $(\bar{\bl Q}(t), \bar{\bl A}(t), \bar{\bl D}(t))$
be given. Thus, there exists a sequence $\{r_n\}_{n=1}^\infty$, such
that as $r_n\rightarrow \infty$, we have
\begin{equation*}
({{\bl Q}(r_nt)\over r_nt}, {{\bl A}(r_nt)\over r_nt}, {{\bl
D}(r_nt)\over r_nt})\rightarrow(\bar{\bl Q}(t), \bar{\bl A}(t),
\bar{\bl D}(t))
\end{equation*}
where the convergence is interpreted as uniformly on compact sets
(u.o.c). Consider the Lyapunov function $L(\bar{\bl
Q}(t))=\max_{i\in\ml V} \bar{Q}_i(t)$, and let $t>0$ be given. Now
it is sufficient to argue that if $\bl a\in \ml A_{\text{LQF}}$, we
have $\dot{L}(\bar{\bl Q}(t))\leq 0$, and therefore stability
follows from Lemma \ref{lem_fluid_stable}.

At time $t$, denote $\ml V_0$ as the set of links with the longest
queues in the fluid limit, i.e., with $\bar{Q}_i(t)=\max_{j\in \ml
V}\bar{Q}_j(t)$. Thus, since the function $\bar{\bl Q}(t)$ is
absolutely continuous, there exists $\epsilon>0$ and $\delta>0$ such
that $\forall \tau\in(t-\delta, t+\delta)$, we have
$\bar{Q}_i(\tau)-\bar{Q}_j(\tau)\geq \epsilon$ for any $i\in\ml V_0,
j\in\ml V_0^c$. Thus, in the original network, for sufficiently
large $n$ we have
\begin{equation*}
{Q}_i(r_n\tau)-{Q}_j(r_n\tau)\geq r_n\epsilon\geq 1, \forall
(r_n\tau)\in(r_n(t-\delta), r_n(t+\delta))
\end{equation*}
for any $i\in\ml V_0, j\in\ml V_0^c$. Therefore, in the original
network during the time interval $(r_n(t-\delta), r_n(t+\delta))$,
none of the links in the set $\ml V_0$ has empty queue. Further,
according to LQF, the links in $\ml V_0$ will always be considered
by LQF before any link in $\ml V_0^c$. Thus, for any link $i\in\ml
V_0$ and any $\tau\in(r_n(t-\delta), r_n(t+\delta))$, if none of
link $i$'s neighbor in $\ml V_0$ are scheduled, link $i$ will be
scheduled for transmission, i.e.,
\begin{equation}
\Delta D_i(\tau)+\sum_{j\in \ml N_i}\Delta D_j(\tau){\bf 1}_{\{j\in
\ml V_0\}}\geq 1
\end{equation}
where $\Delta D(\tau)=D(\tau)-D(\tau-1)$. After summing over the
time interval $(r_n(t-\delta), r_n(t+\delta))$ we conclude that
\begin{eqnarray*}
&&D_i(r_n(t+\delta))+\sum_{j\in\ml
N_i}D_j(r_n(t+\delta)){\bf 1}_{\{j\in \ml V_0\}}\\
&\geq&D_i(r_n(t-\delta))+\sum_{j\in\ml N_i} D_j(r_n(t-\delta))){\bf
1}_{\{j\in \ml V_0\}}+ 2r_n\delta
\end{eqnarray*}
from which we conclude that $\forall i$,
\begin{equation}\label{eqn_sum_dt}
\dot{\bar{D}}_i(t)+\sum_{j\in\ml N_i}\dot{\bar{D}}_j(t){\bf
1}_{\{j\in \ml V_0\}}\geq 1
\end{equation}
Since $\bl a\in \ml A_{\text{LQF}}$, there exists $P\in\ml P$ such
that $\|P\bl a\|_\infty<1$. Denote $i^\star$ as the link in $\ml
V_0$ with the lowest priority, we have
\begin{eqnarray}
a_{i^\star}+\sum_{j\in\ml N_{i^\star}}a_j{\bf 1}_{\{j\in \ml
V_0\}}&\stackrel{(a)}{=}&a_{i^\star}+\sum_{j\in\ml
N_{i^\star}}a_j{\bf 1}_{\{j\in \ml
V_0\}}{\bf 1}_{\{p_{i^\star}>p_j\}}\nonumber\\
&\leq& a_{i^\star}+\sum_{j\in\ml N_{i^\star}}a_j{\bf
1}_{\{p_{i^\star}>p_j\}}\leq 1\label{eqn_diff_qi}
\end{eqnarray}
where $(a)$ is because link $i^\star$ has the lowest priority in
$\ml V_0$. Thus, we have
\begin{eqnarray*}
&&\dot{\bar{Q}}_{i^\star}(t)+\sum_{j\in\ml
N_{i^\star}}\dot{\bar{Q}}_j(t){\bf 1}_{\{j\in
\ml V_0\}}\\
&\stackrel{(a)}{=}&a_{i^\star}+\sum_{j\in\ml
N_{i^\star}}a_j-(\dot{\bar{D}}_{i^\star}(t)+\sum_{j\in\ml
N_i}\dot{\bar{D}}_j(t){\bf 1}_{\{j\in \ml V_0\}})\\
&\stackrel{(b)}{\leq}&a_{i^\star}+\sum_{j\in\ml
N_{i^\star}}a_j-1\leq 0
\end{eqnarray*}
where $(a)$ is because of SLLN and that link $i^\star$ has the
lowest priority in $\ml V_0$, and $(b)$ is because of
(\ref{eqn_sum_dt}). Thus, (\ref{eqn_diff_qi}) shows that
$\dot{\bar{Q}}_{i^\star}(t)\leq 0$. Note that for any regular $t>0$,
we have $\dot{\bar{Q}}_j(t)=\dot{\bar{Q}}_{i^\star}(t)$ for all
$j\in \ml V_0$. Therefore, we conclude that $\dot{L}(\bar{\bl
Q}(t))=\dot{\bar{Q}}_{i^\star}(t)\leq 0$ and the theorem follows by
applying Lemma \ref{lem_fluid_stable}.
\end{IEEEproof}

We next show the tightness of the stability region $\ml
A_{\text{LQF}}$ in some scenarios. We will use the example proposed
in \cite{joo08}. Consider the 6-ring network in Fig.
\ref{fig:sample}, and let the arrival rate be $\bl
a=(1/3+\epsilon)\bl 1$, where $\bl 1$ is the all-ones vector. Define
a periodic arrival process as follows: At time slot $3k+1$, one
packet arrives at link 1 and 4, and at time slot $3k+2$, one packet
arrives at link 2 and 5, and at time slot $3k+3$, one packet arrives
at link 3 and 6. Additionally, in each time slot, with probability
$\epsilon$, one packet arrives at each and every link in the
network. Note that LQF will always schedule $(1,4)$, $(2,5)$ and
$(3, 6)$ and all the queues grow unbounded with rate $\epsilon$.
Therefore, for this network there is an arrival rate vector which is
arbitrarily close to $\ml A_{\text{LQF}}$, which can not be
stabilized by LQF scheduling.

Note that even if the close-form stability region $\ml
A_{\text{LQF}}$ is given, testing whether $\bl a\in\ml
A_{\text{LQF}}$ is still a nontrivial problem since one needs to
consider $n!$ priorities. However, we now show that we can test
whether $\bl a\in\ml A_{\text{LQF}}$ efficiently. In fact, the
following algorithm Test-Feasibility, can achieve this with only
linear complexity.

\begin{algorithm}
\caption{Test-Feasibility $(\ml{G},{\bl a})$}
\begin{algorithmic}
\FOR{$k=n$ to $1$}

\STATE $s_k=\arg\min_{i\in \ml{V}} \{{a}_i+\sum_{j\in
\ml{N}_i}{a}_j\}$;

\IF{$(a_{s_k}+\sum_{j\in \ml{N}_{s_k}}{a}_j>1)$}

\RETURN FALSE;

\ELSE \STATE Remove link $s_k$ and its incident edges from $\ml{G}$;

\ENDIF

\ENDFOR

\RETURN TRUE;
\end{algorithmic}
\end{algorithm}

We finally conclude this section with the following theorem.
\begin{theorem}
Algorithm TF will return TRUE if and only if  $\bl a\in\ml
A_{\text{LQF}}$.
\end{theorem}

Due to space limit, we only describe the intuition behind the proof.
The formal proof is similar to the proof of Theorem 3 in
\cite{li08}. Essentially, if any link $i$ satisfies
$a_{i}+\sum_{j\in \ml{N}_{i}}{a}_j\leq 1$, we say that link $i$ is
worst-case stable, i.e., for any priority vector $\bl p$, we have
$(P\bl a)_i\leq 1$. Thus, if we assign $i$ the lowest priority,
$(P\bl a)_i=a_{i}+\sum_{j\in \ml{N}_{i}}{a}_j\leq 1$, and for any
$j\neq i$, $(P\bl a)_j$ can only get smaller since $i$ now has the
lowest priority. Thus, we are not losing optimality by reassigning
the lowest priority to a worst-case stable link $i$. The proof
follows by repeating the above arguments.

\section{Static Priority Scheduling}
\label{sec_sp}

The stability region of LQF scheduling, $\ml A_{\text{LQF}}$, can be
improved by using SP scheduling, as this section shows. In this
section we consider SP scheduling with multiple priorities, which
are parameterized by $\{\bl p^{(k)}, \bl a^{(k)},
\theta^{(k)}\}_{k=1}^K$, where $\sum_{k=1}^K \bl a^{(k)}=\bl a$ and
$\sum_{k=1}^K \theta^{(k)}=1$. The scheduling algorithm is described
as follows: We divide time slots into blocks where each block with
length $T$ consists of $k$ sub-blocks, such that the $k$-th block
has a length of $\theta^{(k)}T$. Further, each link $i\in\ml V$ has
$K$ sub-queues where each sub-queue $k$ has arrival rate
$a_i^{(k)}$, so that $\sum_{k=1}^K a_i^{(k)}=a_i$. Note that this
can be achieved by filtering the arrival processes probabilistically
into $K$ sub-queues. During the scheduling in each $k$-th time
block, only the sub-queues indexed by $k$ are allowed for
transmission, which follows the order as specified by priority
vector $\bl p^{(k)}$. We have the following theorem about the
stability region.

\begin{theorem}
The network is rate stable under the SP scheduling as described
above if
\begin{eqnarray}\label{eqn_sp_stable}
\|P^{(k)}\bl a^{(k)}\|_\infty<\theta^{(k)}, \forall 1\leq k\leq K
\end{eqnarray}
where $P^{(k)}$ is the incidence matrix associated with $\bl
p^{(k)}$.
\end{theorem}
\begin{IEEEproof}
Let $1\leq k\leq K$ be given, and consider any fluid limit
$(\bar{\bl Q}^{(k)}(t), \bar{\bl A}^{(k)}(t), \bar{\bl D}^{(k)}(t))$
with a converging sequence $\{(\bar{\bl Q}^{(k)r_n}(t), \bar{\bl
A}^{(k)r_n}(t), \bar{\bl D}^{(k)r_n}(t))\}_{n=1}^\infty$. We will
argue that every sub-queue is stable, i.e., $\bar{\bl
Q}^{(k)}(t)=\bl 0$ for all $t\geq 0$ if $\bar{\bl Q}^{(k)}(t)=\bl
0$, and therefore stability follows from Lemma
\ref{lem_fluid_stable}. For the $k$-th sub-queues, define Lyapunov
function $L(\bar{\bl Q}^{(k)}(t))={1\over 2}\|\bar{\bl
Q}^{(k)}(t)\|_2^2$, and consider the link $i_1$ with the highest
priority according to $\bl p^{(k)}$. Suppose that
$\bar{Q}^{(k)}_{i_1}(t)\geq \epsilon>0$ at time $t>0$. Then there
exists $\delta>0$ such that $\bar{Q}^{(k)}_{i_1}(\tau)\geq
\epsilon/2$ for $\tau\in(t-\delta, t+\delta)$. Thus, for
sufficiently large $n$, we have $Q^{(k)}_{i_1}(r_n\tau)/r_n\tau\geq
\epsilon/4$ for $(r_n\tau)\in(r_n (t-\delta), r_n(t+\delta))$, i.e.,
if we choose $n$ large enough, $Q^{(k)}_{i_1}(\tau)$ is never empty
during the time interval $(r_n (t-\delta), r_n(t+\delta))$. Thus,
according to SP scheduling, there is $\theta^{(k)}T$ packet
departures from $Q^{(k)}_{i_1}(\tau)$ in every time block with
length $T$, and we conclude that $\diff \bar{D}_i^{(k)}(t)=
\theta^{(k)}$, and therefore
\begin{equation*}
\diff{1\over 2}{\bar{Q}}_{i_1}^{(k)2}(t)
=\bar{Q}_{i_1}^{(k)}(t)\dot{\bar{Q}}_{i_1}^{(k)}(t)=\bar{Q}_{i_1}^{(k)}(t)(a^{(k)}_{i_1}-\theta^{(k)})\stackrel{(a)}{\leq}
0
\end{equation*}
where $(a)$ is because $a^{(k)}_{i_1}=(P^{(k)}\bl a^{(k)})_{i_1}\leq
\|P^{(k)}\bl a^{(k)}\|_\infty\leq \theta^{(k)}$ since $i_1$ has the
highest priority according to $\bl p^{(k)}$. Thus, we conclude that
${\bar{Q}}_{i_1}^{(k)}(t)=0$ for all $t\geq 0$.

Now suppose this is true for the links $i_1, i_2, \ldots, i_l$,
i.e., the $l$ links with the highest priorities. Now consider link
$i_{l+1}$, and suppose that ${\bar{Q}}_{i_{l+1}}^{(k)}(t)>0$ at some
time $t>0$. Using similar arguments, we have
\begin{eqnarray*}
\diff{1\over
2}{\bar{Q}}_{i_{l+1}}^{(k)2}(t)&=&\bar{Q}_{i_{l+1}}^{(k)}(t)\dot{\bar{Q}}_{i_{l+1}}^{(k)}(t)\\
&\stackrel{(a)}{=}&\bar{Q}_{i_{l+1}}^{(k)}(t)(\dot{\bar{Q}}_{i_{l+1}}^{(k)}(t)+\sum_{j\in
\ml
N_i}\dot{\bar{Q}}_{i_j}^{(k)}(t){\bf 1}_{\{p_i^{(k)}>p^{(k)}_j\}})\\
&=&\bar{Q}_{i_{l+1}}^{(k)}(t)(a^{(k)}_{i_{l+1}}+\sum_{j\in\ml
N_i}a^{(k)}_j{\bf 1}_{\{p_i^{(k)}>p^{(k)}_j\}}-\theta^{(k)})\\
&\leq& 0
\end{eqnarray*}
where $(a)$ is because $\bar{Q}^{(k)}_{i_j}(t)=0$ for links $1\leq
j\leq l$. Thus, induction shows that
${\bar{Q}}_{i_{l+1}}^{(k)}(t)=0$ for all $t>0$, and therefore, the
network is rate stable.
\end{IEEEproof}

We have the following corollary, which says that it is sufficient to
consider less than $n+1$ priorities to achieve the optimal stability
region $\ml A_{\max}$ \cite{tassiulas92}, which is the convex hull
of the set of independent sets.

\begin{corollary}
Let $\bl a\in \ml A_{\max}$ be given. For $K=n+1$, there exists a SP
scheduling with parameters $(\bl p^{(k)}, \bl a^{(k)},
\theta^{(k)})_{k=1}^K$ such that $\bl a$ is stable.
\end{corollary}
\begin{IEEEproof}
Note that this is similar to the statement in \cite{li09}. We
briefly provide the proof here for completeness. Since $\bl a\in \ml
A_{\max}$, from Carath\'eodory Theorem, we can express $\bl a$ as a
convex combination of at most $n+1$ independent sets, i.e., $\bl
a=\sum_{k=1}^{n+1}\theta^{(k)}\bl m^{(k)}$, where $\bl m^{(k)}$ is
an $n\times 1$ indicator vector representing an independent set,
i.e., $m^{(k)}_i=1$ if link $i$ is included in the set, otherwise
$m^{(k)}_i=0$. Now define $\bl a^{(k)}=\theta^{(k)}\bl m^{(k)}$, and
choose priority $\bl p^{(k)}$ such that the links in the independent
set $\bl m^{(k)}$ have locally highest priority. Thus, $P^{(k)}$ is
a diagonal matrix and $P^{(k)}\bl a^{(k)}=\bl a^{(k)}$. With this
choice of $(\bl p^{(k)}, \bl a^{(k)}, \theta^{(k)})_{k=1}^K$, it is
easily seen that (\ref{eqn_sp_stable}) is satisfied and the network
is stable.
\end{IEEEproof}

Although the lower bound stability region for SP scheduling is
known, the assignment of parameters $(\bl p^{(k)}, \bl a^{(k)},
\theta^{(k)})_{k=1}^K$ is not trivial. For the special case with
$K=1$ and $K=n+1$, we have shown the optimal priority assignment
algorithms \cite{li09}. For the cases $2\leq k\leq K$, it is, in
general, hard to assign the parameters optimally. In the following
we consider a special case, where we assume two priorities $\bl
p^{(1)}, \bl p^{(2)}$ and $\theta^{(1)}=\theta^{(2)}=1/2$, for the
ease of implementation. In this case, we can obtain the priorities
by solving the following problem:
\begin{eqnarray*}
\textsf{OPT: }\min_{P^{(1)}, P^{(2)}, \bl x}&&\max(\|P^{(1)}\bl x\|_\infty, \|P^{(2)}(\bl a-\bl x)\|_\infty)\\
\textrm{subject to}&& \bl 0\preceq \bl x\preceq \bl a\\
&&P^{(1)}, P^{(2)}\in \ml P
\end{eqnarray*}
Note that even in this special case, an optimal solution is hard to
get since the problem is non-convex. We next propose a heuristic
algorithm (Alg. \ref{alg_em}) to solve the above problem, which is
related to the well-known EM algorithm in the literature.

\begin{algorithm}
\caption{EM$(\ml P, \bl a)$}
\begin{algorithmic}\label{alg_em}
\WHILE{(1)}

\STATE 1) \textsf{E-Step: }Choose priorities $(P^{(1)\star},
P^{(2)\star})$ by solving
\begin{eqnarray*}
&&\min_{P^{(1)}, P^{(2)}\in \ml P}\max(\|P^{(1)}\bl
x^\star\|_\infty,
\|P^{(2)}(\bl a-\bl x^\star)\|_\infty)\\
&=&\max(\min_{P^{(1)}\in \ml P}\|P^{(1)}\bl x^\star\|_\infty,
\min_{P^{(2)}\in \ml P}\|P^{(2)}(\bl a-\bl x^\star)\|_\infty)
\end{eqnarray*}
\STATE 2) \textsf{M-Step: } Choose arrival rates $(\bl x^\star, \bl
a-\bl x^\star)$ by solving
\begin{eqnarray*}
&\min_{\bl x, t}& t\\
&\textrm{subject to} & P^{(1)\star}\bl x\preceq (t/2)\bl 1\\
&& P^{(2)\star}(\bl a-\bl x)\preceq (t/2)\bl 1\\
&& \bl 0\preceq \bl x\preceq \bl a
\end{eqnarray*}

\RETURN if the sequence of $t$'s has converged

\ENDWHILE

\end{algorithmic}
\end{algorithm}

After each iteration, the objective function value in \textsf{OPT}
gets smaller, and therefore converges to a local optimal assignment.
If the limit of $t$'s is less than 1, $\bl a$ is stable using the 2
priority vectors. We will test the performance of the EM algorithm
in the next section.

\section{Simulation Results}
\label{sec_simulation}

In this section we demonstrate the performance of LQF and SP
scheduling through simulation. Since the performance of LQF and SP
scheduling is dependent on the arrival processes as well as the
network topology, an exhaustive search of arrival processes and the
networks is clearly not possible. Thus, similar to the recent
research \cite{birand10}, where specific graph structures are
explored to demonstrate the performance limit of scheduling, we will
consider special network topologies and arrival processes to
demonstrate the performance limit of LQF and SP scheduling.

\subsection{6-Ring}

\begin{figure}
\centering
\includegraphics[width=3.5in]{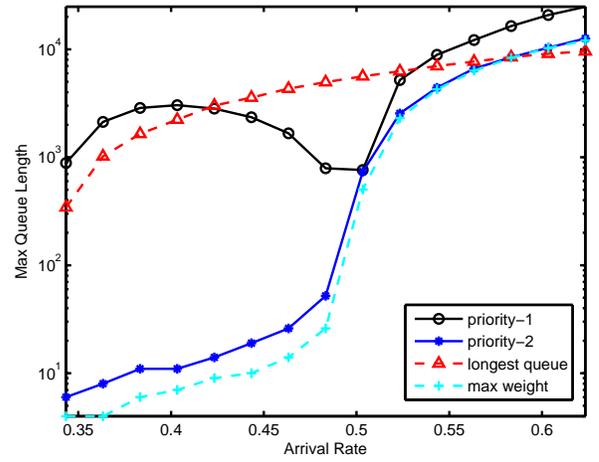}
\caption{The simulation result of a 6-ring network with different
schedulers.} \label{fig_ring_6}
\end{figure}

We first consider the 6-ring network in Fig. \ref{fig:sample}(a) and
the arrival process as described in Section \ref{sec_lqf} ($\bl a$
is uniform). Fig. \ref{fig_ring_6} shows the network stability
result with respect to the uniform arrival rate over a time period
of $10^5$ time slots, where the result is averaged over 10
independent simulations. Note that boundary of the optimal stability
region is at a uniform arrival rate of $0.5$, above which the clique
constraint (i.e., a single edge) is violated. One can observe that
neither the LQF scheduling nor the SP scheduling with a single
priority vector is stable, as the max queue length is large (the
decreasing behavior of SP scheduling with single priority near 0.5
is due to the specific arrival process) However, both the Max-Weight
scheduling \cite{tassiulas92} and SP scheduling with two (optimal)
priority vectors can stabilize the network.

\subsection{Bipartite Graph}

\begin{figure}
\centering
\includegraphics[width=3.5in]{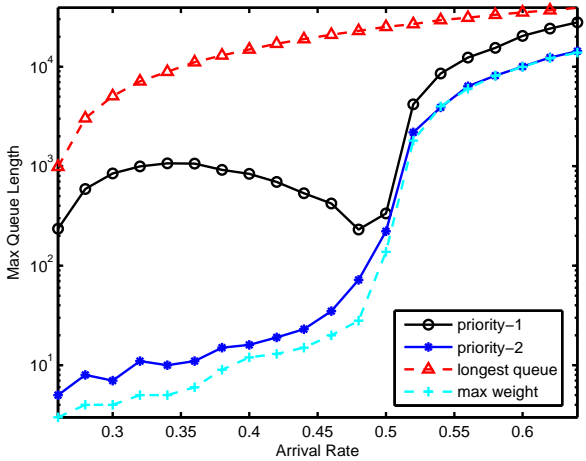}
\caption{The simulation result of a bipartite network with different
schedulers.} \label{fig_bipartite}
\end{figure}

We next consider an incomplete bipartite graph with 8 links as shown
in Fig. \ref{fig:sample}(b). We consider a periodic arrival process
which is similar to the one for the 6-ring network. The stability
result is shown in Fig. \ref{fig_bipartite}. Similarly, one can
observe that the LQF scheduling and SP scheduling with single
priority vector is not stable, whereas both the SP scheduling with
two priorities and the max-weight scheduling can stabilize the
maximum uniform arrival rate, which is 0.5 for this network.

\section{Conclusion}
\label{sec_conclusion}

In this paper we considered two greedy scheduling algorithms in
wireless networks: LQF scheduling and SP scheduling. For LQF
scheduling, we formulated a close-form stability region, which was
shown to be tight in some scenarios. For SP scheduling with multiple
priorities, we also proposed a lower bound stability region, as well
as a heuristic priority assignment algorithm, which is related to
the well-known EM algorithm. The performance gain of the proposed
heuristic algorithm was finally confirmed by simulations.

\appendix

In the appendix we introduce fluid limits \cite{dai00}, which are
used to prove rate stability.

\subsection{Existence of Fluid Limits}

Given the network dynamics which are described by the functions
$({\bl Q}(t), {\bl A}(t), {\bl D}(t))_{t=0}^\infty$, we first extend
the support from $\mathbb{N}$ to $\mathbb{R}_+$ using linear
interpolation. Now, for a fixed sample path $\omega$, define the
fluid scaling of a function $f(t)$ as $f^r(t)={f(rt)/r}$, where $f$
can be ${Q}_i(t), {A}_i(t)$ or ${D}_i(t)$. It can be verified that
these functions are uniformly Lipschitz-continuous, i.e., for any
$t>0$ and $\delta>0$, we have
\begin{eqnarray}
A_i^{r}(t+\delta)-A_i^{r}(t)&\leq&A_{\max}\delta\\
D_i^{r}(t+\delta)-D_i^{r}(t)&\leq&\delta\\
Q_i^{r}(t+\delta)-Q_i^{r}(t)&\leq&A_{\max}\delta
\end{eqnarray}
Thus, these functions are equi-continuous. According to
Arz\'ela-Ascoli Theorem, any sequence of functions
$\{f^{r_n}(t)\}_{n=1}^\infty$ contains a subsequence
$\{f^{r_{n_k}}(t)\}_{k=1}^\infty$, such that such that
\begin{equation}
\lim_{k\rightarrow\infty}\sup_{\tau\in[0,t]}|f^{r_{n_k}}(\tau)-\bar{f}(\tau)|=0
\end{equation}
where $\bar{f}(t)$ is a uniformly continuous (and therefore
differentiable almost everywhere) function. In our example, $f(t)$
can be ${Q}_i(t), {A}_i(t)$ or ${D}_i(t)$. Define any such limit
$(\bar{\bl Q}(t), \bar{\bl A}(t), \bar{\bl D}(t))$ as a fluid limit.

\subsection{Properties of Fluid Limits}

We have the following properties holds for any fluid limit
\begin{eqnarray}
\bar{A}_i(t)&=&a_it\label{eqn_ai_bar}\\
\diff \bar{Q}(t) &=&0\quad\textrm{ if }\bar{Q}(t)=0\label{eqn_q_bar}
\end{eqnarray}
where (\ref{eqn_ai_bar}) is because of (functional) SLLN, and
(\ref{eqn_q_bar}) is because any regular point $t$ with
$\bar{Q}(t)=0$ achieves local minimum (since $\bar{Q}(t)\geq 0$),
and therefore has zero derivative. We further have the following
lemma which states a sufficient condition about rate stability
\cite{dai00}:
\begin{lemma}\label{lem_fluid_stable}
The network is rate stable if any fluid limit with $\bar{\bl
Q}(0)={\bf 0}$ has $\bar{\bl Q}(t)={\bf 0}, \forall t\geq 0$.
\end{lemma}

\end{document}